\documentclass[twocolumn]{aastex631}
\usepackage{graphicx}
\usepackage{hyperref}
\usepackage{booktabs}
\usepackage{makecell}
\usepackage{amsmath}

\begin{document}

\title{Probing the Nature of High‑Redshift Long GRB 250114A and Its Magnetar Central Engine}
\author{Wen-Yuan Yu}
\affiliation{Guangxi Key Laboratory for Relativistic 
Astrophysics, School of Physical Science and Technology, Guangxi University, Nanning 530004, China}
\author{Hou-Jun L\"{u}}
\altaffiliation{Corresponding author (LHJ) email: lhj@gxu.edu.cn}
\affiliation{Guangxi Key Laboratory for Relativistic 
Astrophysics, School of Physical Science and Technology, Guangxi University, Nanning 530004, China}
\author{Xiao Tian}
\affiliation{Guangxi Key Laboratory for Relativistic 
Astrophysics, School of Physical Science and Technology, Guangxi University, Nanning 530004, China}
\author{Liang-Jun Chen}
\affiliation{Guangxi Key Laboratory for Relativistic 
Astrophysics, School of Physical Science and Technology, Guangxi University, Nanning 530004, China}
\author{En-Wei Liang}
\affiliation{Guangxi Key Laboratory for Relativistic 
Astrophysics, School of Physical Science and Technology, Guangxi University, Nanning 530004, China}

\begin{abstract}
GRB 250114A is a long-duration gamma-ray burst (GRB) which triggered the Swift/BAT with a spectroscopic high-redshift at $z = 4.732$. The light curve of the prompt emission is composed of three distinct emission episodes, which are separated by quiescent gaps ranging from tens to hundreds of seconds. While the X-ray light curve exhibits the canonical X-ray emission which is composed of several power-law segments superposition of a giant X-ray flare. More interestingly, there is still significant X-ray emission during the quiescent time in the prompt emission, suggesting a continuously active central engine whose power fluctuates across the $\gamma$‑ray detectability threshold. In this paper, we propose a magnetar as the central engine of GRB 250114A by fitting the X-ray light curve, and infer a magnetic field strength $B_{\rm p}=13.24^{+1.73}_{-5.84} \, \times10^{15}\ \mathrm{G}$ and an initial spin period $P_{0}=14.31^{+0.93}_{-3.16} \,  \mathrm{ms}$ of magnetar, with a jet correction, fall within a reasonable range. Furthermore, we also compare the prompt emission, X-ray afterglow, $E_{\mathrm p}$–$E_{\gamma,\mathrm{iso}}$, and $\varepsilon-$distribution of GBR 250114A with those of other high-$z$ sample-GRBs, and find no significant statistical differences between them.

\end{abstract}


\section{Introduction}
Gamma-ray bursts (GRBs) are the most violent and energetic explosions in the universe. Phenomenologically, they are generally classified into two types (e.g., long and short) with a division line at the observed duration $T_{90}\sim 2$ s \citep{1993ApJ...413L.101K}. The long GRBs are thought to be from the core-collapse of massive stars \citep{1998Natur.395..670G, 2003ApJ...591L..17S, 2004ApJ...609L...5M, 2006ApJ...645L..21M, 2006Natur.442.1011P, 2006Natur.444.1010Z, 2006ARA&A..44..507W, 2017AdAst2017E...5C, 2018ApJ...862..130L}, while the short GRBs are proposed to be produced by mergers of two compact stellar objects \citep{2005Natur.438..994B, 2005Natur.438..988B, 2005Natur.437..851G, 2005Natur.437..845F, 2010ApJ...708....9F}, such as neutron star-neutron star mergers (NS-NS; \citealt{1986ApJ...308L..43P, 1989Natur.340..126E}), or neutron star-black hole mergers (NS-BH; \citealt{1991AcA....41..257P}). 

Moreover, several peculiar GRBs, called short GRB with extended emission, are also claimed to originate from the merger of two compact stars, including GRB 060614 \citep{2006Natur.444.1044G, 2015NatCo...6.7323Y}, GRB 211227A \citep{2022ApJ...931L..23L, 2023A&A...678A.142F}, GRB 211211A \citep{2022Natur.612..223R, 2022Natur.612..228T, 2022Natur.612..232Y, 2023ApJ...943..146C,2025ApJ...988L..46L}, and GRB 230307A \citep{2023ApJ...954L..29D, 2024Natur.626..737L, 2024Natur.626..742Y, 2024ApJ...962L..27D, 2025NSRev..12E.401S}. Recently, kilonova (KNe) emission has been observed in some of these sources, further supporting their compact binary merger origin \citep{2025JHEAp..45..325Z,2025ApJ...988L..46L}. In addition, statistical studies suggest that such GRBs with extended emission have smaller plateau luminosities than the rest of the long GRB sample \citep{2025JHEAp..4700384L}, which makes the merger scenario more favorable.
On the other hand, the coalescence systems are also the main targets as strong sources of gravitational wave (GW) events \citep{2014ARA&A..52...43B}. Especially, on 17 August 2017, the first direct detection of a GW event (GW170817) associated with short GRB 170817A was achieved by the Advanced Laser Interferometer Gravitational-wave Observatory (aLIGO) and Virgo \citep{2017PhRvL.119p1101A, 2017ApJ...848L..15S, 2017ApJ...848L..14G, 2018NatCo...9..447Z}, and it has confirmed that at least some short GRBs are indeed originate from the merger of compact stars. On the other hand, the detection of both GW170817 and GRB 170817A also opened a new window into studying such a catastrophic event across multiple messengers.

Historically, the first measured redshift for GRB 970508 was $z=0.835$, which confirmd the cosmological origin of GRBs \citep{1997Natur.387..878M}. Over the past few decades, the redshift of an increasing number of GRBs have been measured with a range from $z=0.085$ to $z=9.4$ \citep{1998Natur.395..670G, 2011ApJ...736....7C}. Especially, several high-redshift GRBs are reported, including GRB 080913 at $z=6.7$ \citep{2009ApJ...693.1610G}, GRB 090423 at $z=8.2$ \citep{2009Natur.461.1254T, 2009Natur.461.1258S}, and the most distant candidate with a photometric redshift at $z=9.4$ for GRB 090429B \citep{2011ApJ...736....7C}. On the one hand, the high-redshift of these GRBs has challenged the classification of GRBs \citep{2009ApJ...703.1696Z}. If this is the case, the multiple observational criteria \citep{2009ApJ...703.1696Z}, a new classification of GRBs \citep{2010ApJ...725.1965L} and the three-dimensional of GRB prompt emission \citep{2014MNRAS.442.1922L} have been proposed to study the GRB classification. On the other hand, even though the redshifts of hundreds of GRBs have now been measured, the distribution of these measurements remains heavily weighted towards low‑$z$ events, and the high‑$z$ GRBs are rare and exceptionally valuable. It has been established that high‑$z$ GRBs can be utilized to trace the cosmic star formation rate \citep{2009ApJ...705L.104K, 2011MNRAS.418..500I, 2012ApJ...744...95R}. This enables the precise identification of high-redshift galaxies, facilitating the exploration of their metal and dust content \citep{2009Natur.461.1254T, 2013MNRAS.429.2718S}. Furthermore, they can be used to place limits on the mass of dark matter particles \citep{2013MNRAS.432.3218D} and on the degree of primordial density non-Gaussianity \citep{2012MNRAS.426.2078M}, as well as to gauge the intensity of the local intergalactic radiation field \citep{2010MNRAS.404.1938I}. Additionally, the high‑$z$ GRBs can be employed to constrain fourth-order cosmographic expansions at ($z>1$) \citep{2022A&A...661A..71H}, and provide direct and/or indirect evidence of the first massive, metal-free stars, also known as Population III stars \citep{2012ApJ...760...27W, 2015MNRAS.449.3006M}.

From an observational point of view, the prompt emission of a subset of long GRBs is composed of multiple emission episodes with quiescent times of up to hundreds of seconds, as detected by both the Fermi/Gamma-ray Burst Monitor (GBM) and the Swift/Burst Alert Telescope (BAT; \citealt{1995ApJ...452..145K, 2005MNRAS.357..722L, 2013ApJ...775...67B, 2014ApJ...789..145H, 2018ApJ...862..155L}). \cite{2018ApJ...862..155L} found that certain Fermi/GBM-detected GRBs show multiple emission phases with quiescent intervals, yet the two episodes do not exhibit significant differences in their peak energies ($E_{\rm p}$). Similarly, \cite{2014ApJ...789..145H} reported that some GRBs observed by Swift display multi-stage radiation characteristics. These findings raise important questions: Do these emission episodes originate from different physical processes? Or are they simply manifestations of reactivities of the central engine, with the high-energy component being absent during one of the episodes and only low-energy emission persisting?

Recently, a high‐redshift GRB 250114A was detected by Swift/BAT with a redshift of $z=4.732$ \citep{2025GCN.38927....1K, 2025GCN.38934....1M}. The prompt emission of GRB 250114A consists of a triple-episode structure with two long-duration gaps which are quite similar to those observed precursors by the Burst And Transient Source Experiment (BATSE) \citep{1995ApJ...452..145K} and the Fermi/GBM \citep{2018ApJ...862..155L, 2020PhRvD.102j3014C}, while the X-ray afterglow light curve exhibits a flare phase and a plateau emission. More interestingly, significant X-ray emission was observed during the quiescent time in the prompt emission. Based on the properties of its multiwavelength data presented here, the central engine of GRB 250114A appears to be a magnetar. We present our data reduction and temporal‑spectral analyses of data for GRB 250114 in Section~\ref{sec:data}. In Section~\ref{sec:comparison}, we compare the prompt emission and X-ray afterglow of GRB 250114A with those of other high‑$z$ (e.g., $z>4.5$) GRBs. The magnetar‑engine fitting procedure and a discussion of the physical implications are presented in Section~\ref{sec:model}. Conclusions and discussion are
drawn in Section~\ref{sec:conclusion} and Section~\ref{sec:Discussion}, respectively. Throughout the paper, we adopt a concordance cosmology with $\Omega_M=0.30$, $\Omega_\Lambda=0.70$, and $H_0=70\,\mathrm{km\,s^{-1}\,Mpc^{-1}}$.

\section{Data reduction and analysis} \label{sec:data}
\subsection{Swift/BAT Observations} 
\begin{figure}[htbp]
\centering
\includegraphics[width=0.45\textwidth]{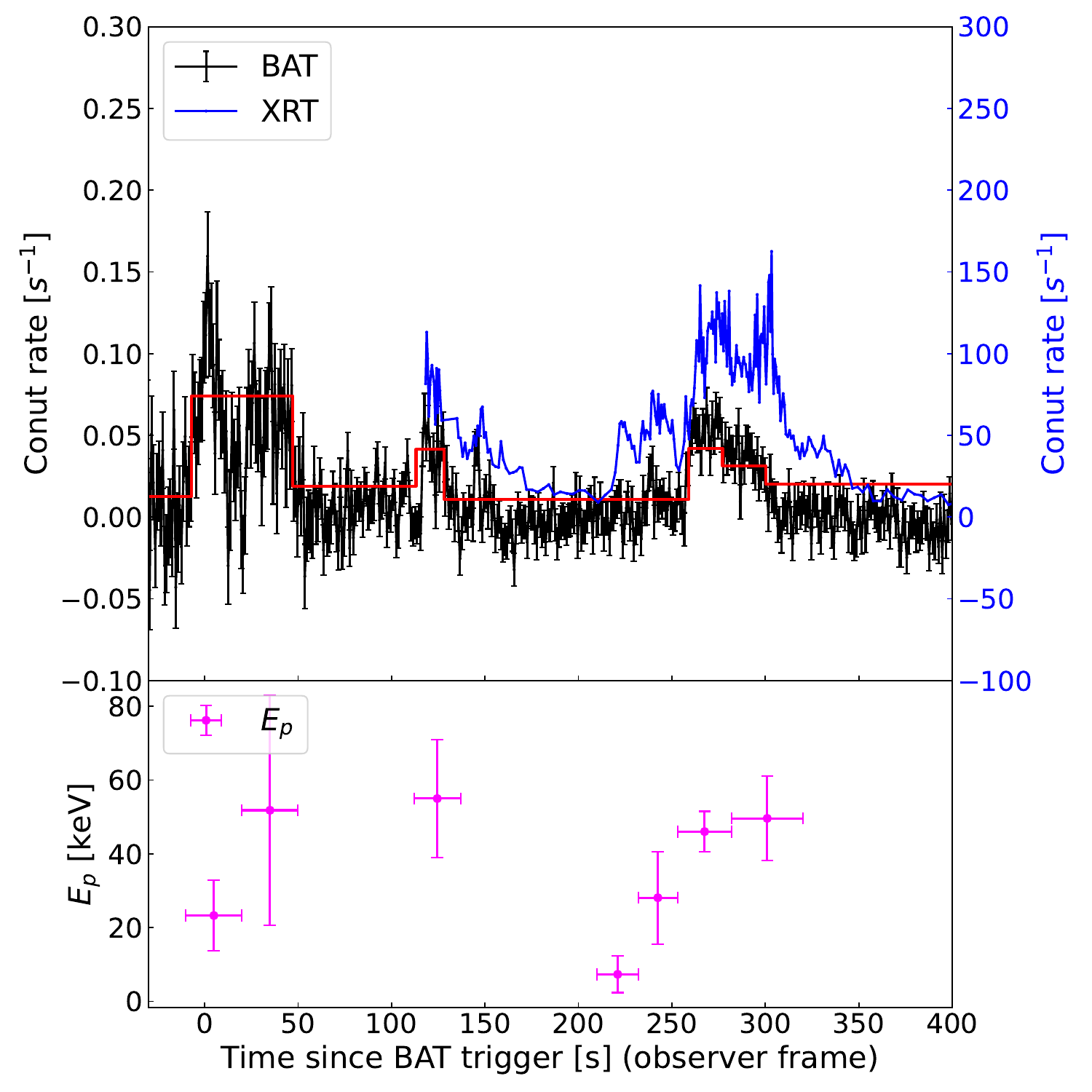}
\caption{Light curves of prompt emission (black line in top panel) observed by BAT and X-ray emission (blue line in top panel), as well as $E_{\rm p}$ evolution (bottom panel) of GRB 250114A. The red step line is Bayesian block analysis for BAT light curve.
\label{fig1}}
\end{figure}

 GRB 250114A initially triggered the Swift/BAT on 14 January 2025 at 07:33:19 UT \citep{2025GCN.38927....1K}. We downloaded the BAT data from the Swift website\footnote {\url{https://www.swift.ac.uk/swift_portal/getobject.php?name=GRB+250114A}}, and use the standard HEASOFT v6.34 to process the BAT data. The BAT light curve in 15--350 keV is extracted with fixed 1 s time-bin by using \texttt{batbinevt} and employing the standard mask-weighting technique. The background is extracted using two time intervals before and after the burst, and then model the background as Poisson noise which is the standard back ground model for prompt emission in BAT events. The duration of the prompt emission can be estimated as $ T_{\mathrm{90, BAT}} = 294 \pm 13$ s in the 15-350 keV. Temporal structure analysis via the Bayesian Blocks algorithm \citep{1998ApJ...504..405S} can identify three distinct emission episodes (see Figure~\ref{fig1}), named as episode-\uppercase\expandafter{\romannumeral1}, \uppercase\expandafter{\romannumeral2} and \uppercase\expandafter{\romannumeral3}, with the false-positive probability threshold set to $p_0=0.05$ by using the \texttt{astropy.stats.bayesian\_blocks} implementation. The analysis was performed on the BAT light curve in the 15–350 keV energy band objectively to identify significant temporal changes in the count rate, which serves to define the boundaries of the emission episodes for subsequent temporal analysis.

 We also extract the time-averaged spectrum of the prompt emission observed by BAT, and several spectral models can be selected to test the spectral fitting of the burst, such as power law (PL), cutoff power law (CPL), Band function (Band), and Blackbody (BB), as well as combinations of any two models. It is found that the CPL model is the best one to adequately describe the observed data by comparing the goodness of different model fits to invoke the Bayesian information criteria (BIC). One has photon index $\Gamma = 0.34 \pm 0.63$ and $E_{\mathrm{p}} = 51\pm 6$ keV. Then, we also perform the time-resolved spectrum on each episode to present the spectral evolution. It is found that the episode-\uppercase\expandafter{\romannumeral1} includes two pulses, and the spectrum of each pulse can be fitted well by CPL model (see Table \ref{table1}). One needs to note that there is still significant X-ray emission during episode-\uppercase\expandafter{\romannumeral2}, episode-\uppercase\expandafter{\romannumeral3}, and the quiescent time between them. By jointing to the BAT and X-ray emissions during  episode-\uppercase\expandafter{\romannumeral2} and episode-\uppercase\expandafter{\romannumeral3}, we also find that the CPL model is the best one to fit the time-resolved spectra, the fitting results are shown in Figure~\ref{fig1} and Table \ref{table1}.

\subsection{Swift/XRT and Follow-up Observations} 
\begin{figure}[htbp]
\centering
\begin{minipage}[b]{0.49\textwidth} 
  \centering
  \includegraphics[width=\linewidth]{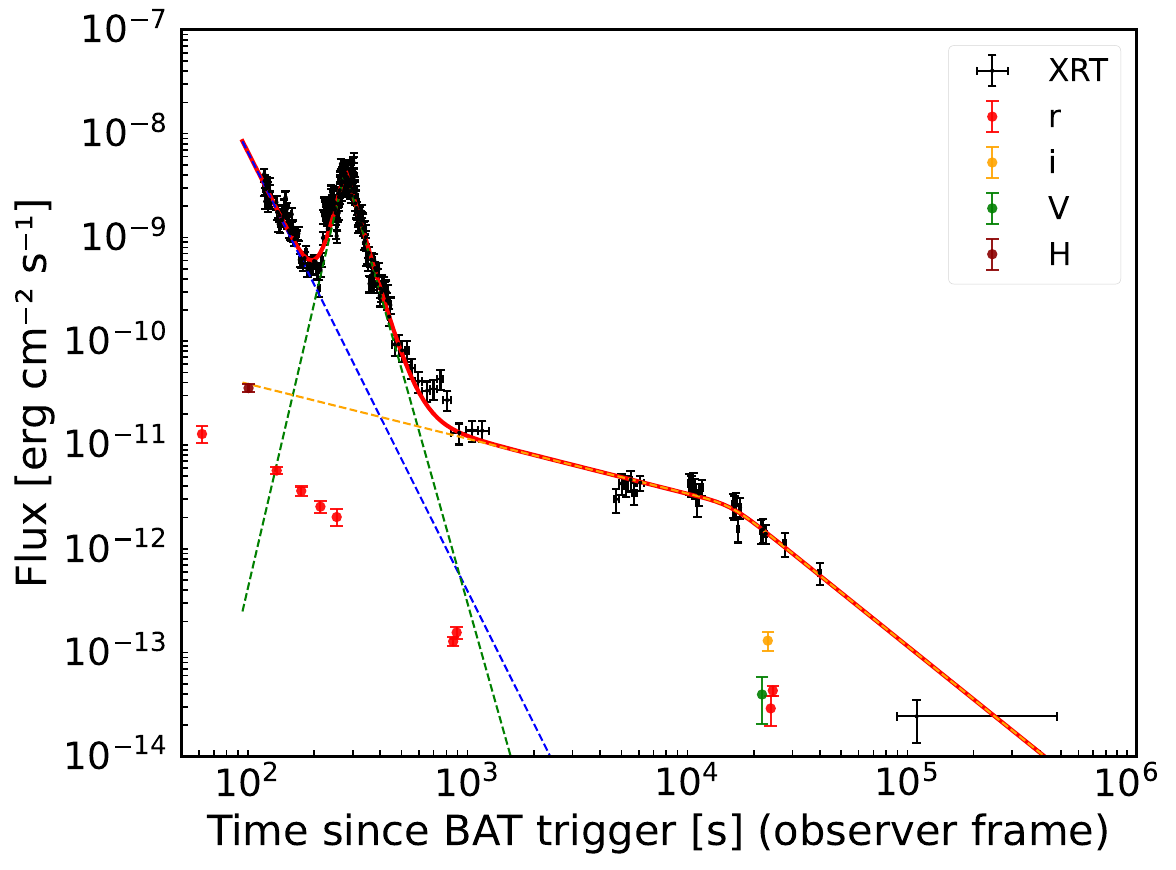} 
\end{minipage}
\hfill
\begin{minipage}[b]{0.45\textwidth}
  \centering
  \includegraphics[width=\linewidth]{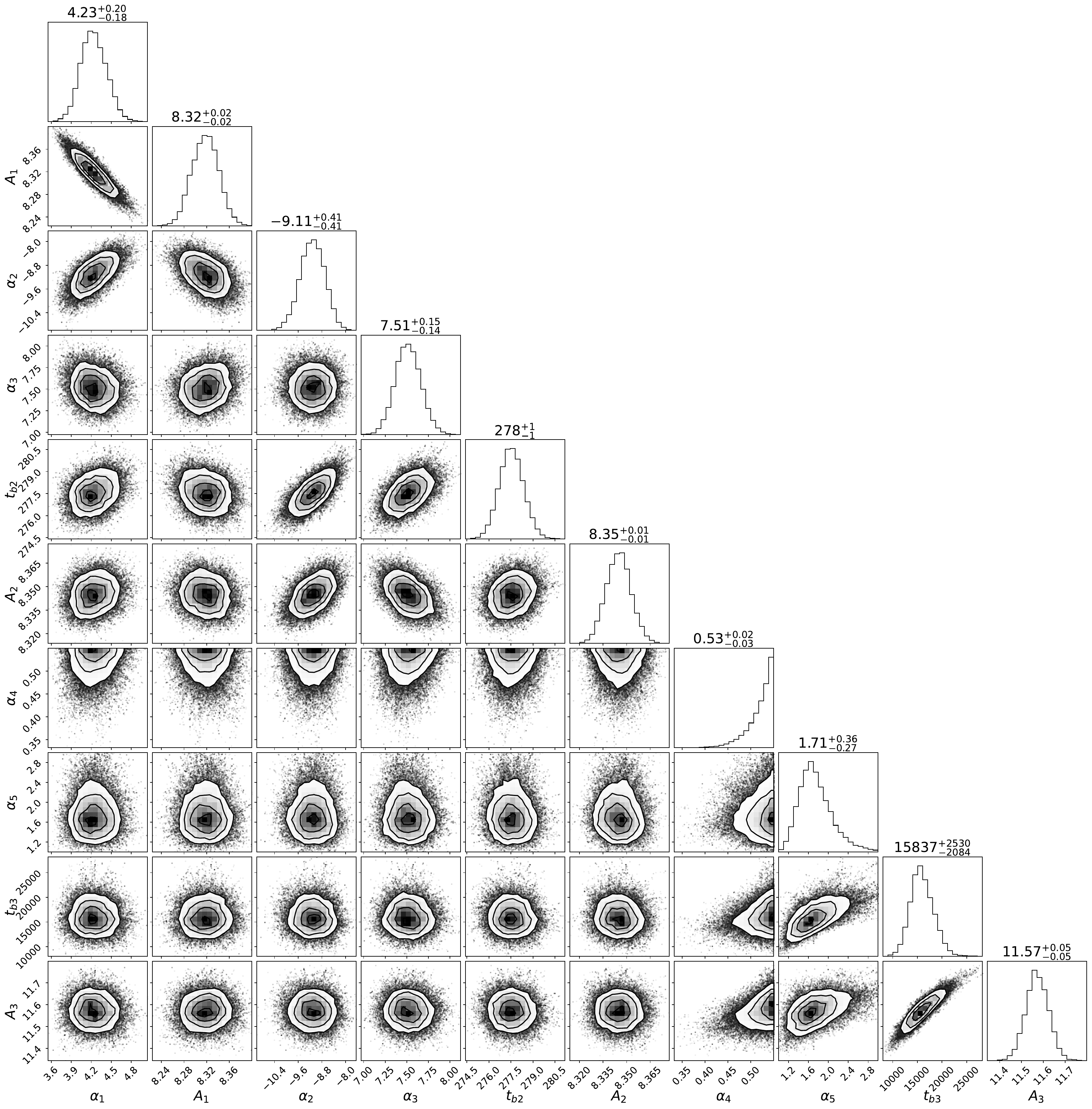}

\end{minipage}
\caption{Left panel: XRT (black) light curve of GRB 250114A. The red solid line represents the total component fitting line. The blue, green, and orange dashed lines are the best fits for the power-law, flare and plateau, respectively. The red, yellow, green and darkred data points are multiband optical data collected from GCN reports. Right panel: corner plots and parameter constraints for the fitting parameters of the XRT light curve of GRB 250114A.}
\label{fig2}
\end{figure}

The Swift X-Ray Telescope (XRT) began observing the field at 117 s after the BAT trigger \citep{2025GCN.38945....1D}, and an optical counterpart was discovered by the Ultraviolet/Optical Telescope (UVOT; \citealt{2025GCN.38927....1K}). Subsequently, spectroscopic observations by the Very Large Telescope (VLT)/X-shooter derived a redshift for this GRB is $z=$4.732 \citep{2025GCN.38934....1M}. Moreover, the optical afterglow follow-up observations are also reported by Gamma-ray Coordinates Network (GCN), and we collated the observed optical data from GCN reports, shown in the left panel of Figure \ref{fig2} \citep{2025GCN.38929....1A, 2025GCN.38946....1A, 2025GCN.38933....1B, 2025GCN.38948....1G, 2025GCN.38930....1M,  2025GCN.38943....1P}.

We downloaded the XRT data of GRB 250114A from the Swift archive\footnote{https://www.swift.ac.uk/xrt$\underline{~}$curves/01281241/}. Figure \ref{fig2} shows the X-ray light curve of GRB 250114A in the energy range of 0.3-10 keV. The X-ray light curve exhibits the canonical X-ray emission \citep{2006Natur.444.1010Z, 2006ApJ...642..389N} that seems to be composed of several power-law segments: an initial steep decay smoothly connected with a plateau phase and followed by a normal decay segment, and a giant X-ray flare is
superimposed on the steep decay segment. Then, we perform an empirical fitting to the X-ray light curve with a power-law $+$ two smooth broken power-law models, it can be expressed as \citep{2007ApJ...670..565L, 2025ApJ...982...19T},
\begin{equation}
    F_{1} = F_{0,1} \left( \frac{t}{t_{1}} \right)^{-\alpha_{1}},
\end{equation}
\begin{equation}
    F_{2} = F_{0,2} \left[ \left( \frac{t}{t_{\mathrm{b},2}} \right)^{\omega_{2}\alpha_{2}} + \left( \frac{t}{t_{\mathrm{b},2}} \right)^{\omega_{2}\alpha_{3}} \right]^{-1/\omega_{2}},
\end{equation}
\begin{equation}
    F_{3} = F_{0,3} \left[ \left( \frac{t}{t_{\mathrm{b},3}} \right)^{\omega_{3}\alpha_{4}} + \left( \frac{t}{t_{\mathrm{b},3}} \right)^{\omega_{3}\alpha_{5}} \right]^{-1/\omega_{3}}.
\end{equation}
The total X-ray light curve can be fitting by the superposition of those three components,
\begin{equation}
    F = F_1 + F_2 + F_3.
\end{equation}
Here, $t_1$ is the starting time of X-ray observations, $t_{\rm b,2}$ and $t_{\rm b,3}$ are the two break times for the flare and plateau emissions. $\alpha_1$ is the initial power-law decay index, $\alpha_2$ and $\alpha_3$ are the rise and decay index before and after $t_{\rm b,2}$, $\alpha_4$ and $\alpha_5$ are the decay index before and after $t_{\rm b,3}$. $\omega_2$ and $\omega_3$ describe the sharpness of the break at $t_{\rm b,2}$ and $t_{\rm b,3}$, and fixed with $\omega_2=10$ and $\omega_3 = 7$.

One can adopt a Markov Chain Monte Carlo (MCMC) method to fit the light curve by invoking above one power-law $+$ two smooth broken power-law model. The first segment is a power-law decay with the temporal index $\alpha_1 = 4.22^{+0.19}_{-0.18}$, which is consistent with the tail emission of prompt $\gamma -$ray from curvature effect \citep{2000ApJ...541L..51K, 2007ApJ...670..565L, 2015ApJ...808...33U}. The flare exhibits a fast rise phase followed by a rapid decay phase with the temporal index $\alpha_2 = -9.14\pm 0.38$ and $\alpha_3 = 7.52^{+0.15}_{-0.14}$, and the break time is $t_{\rm b,2}= 277.50^{+0.80}_{-0.78}$ s. In the plateau phase, the temporal decay index before and after the break time $t_{\rm b,3}= 15860^{+2532}_{-2344}$ s are $\alpha_4 = 0.53^{+0.02}_{-0.03}$ and $\alpha_5 = 1.70^{+0.38}_{-0.28}$, respectively. The peak fluxes at the break time of the flare and plateau are $F_{\rm b,2} = (4.33 \pm 0.08) \times 10^{-9}$ erg cm$^{-2}$ s$^{-1}$ and $F_{\rm b,3} = (2.48 \pm 0.30) \times 10^{-12}$ erg cm$^{-2}$ s$^{-1}$, respectively. The fitting results of X-ray light curve for GRB 250114 are shown in Figure \ref{fig2} and Table \ref{table2}.

\subsection{Overlapping between $\gamma-$ray and X-ray emissions} 
More interestingly, it is found that the X-ray flare emission of GRB 250114A was simultaneously observed by BAT in episode- \uppercase\expandafter{\romannumeral2} and episode-\uppercase\expandafter{\romannumeral3} by comparing the light curves of X-ray and $\gamma-$ray. Then, we analyzed the joint spectral fit for the two emission episodes in the overlapping portion of the XRT and BAT data. For the episode- \uppercase\expandafter{\romannumeral2}, the time-averaged spectrum from 112 s to 137 s is best fit by a cutoff power-law model with $E_\mathrm{p} = {55 \pm 16}$ keV. For the episode-\uppercase\expandafter{\romannumeral3}, we temporally sliced it into four slices based on the shape of the light curve, namely, (210-232) s, (232-253) s, (253-282) s, (282-320) s. All the time-averaged spectral of those slices are best fit by a cutoff power-law model, and the values of $E_\mathrm{p}$ are presented in the Table \ref{table1}\footnote{One needs to note that the spectral fitting results of time intervals (112–137)s, (210–232)s, and (253–282)s are derived from joint XRT+BAT+optical (see 
\citealt{2024ApJ...977..197F}.)} It is found that the spectral evolution is from soft to hard in episode-\uppercase\expandafter{\romannumeral3}.

On the other hand, there is a quiescent period between episode- \uppercase\expandafter{\romannumeral2} and episode-\uppercase\expandafter{\romannumeral3} in the prompt emission, but the so-called quiescent period is not completely quiet as seen in the X-ray light curve from XRT (Figure \ref{fig1}). It is notable that there is still significant X-ray emission during the quiescent period, suggesting that the central engine may be still active, with this phase lasting hundreds of seconds.


\section{Comparison With Other High-Redshift GRBs} \label{sec:comparison}
\begin{figure}[htbp]
\centering
\includegraphics[width=0.6\textwidth]{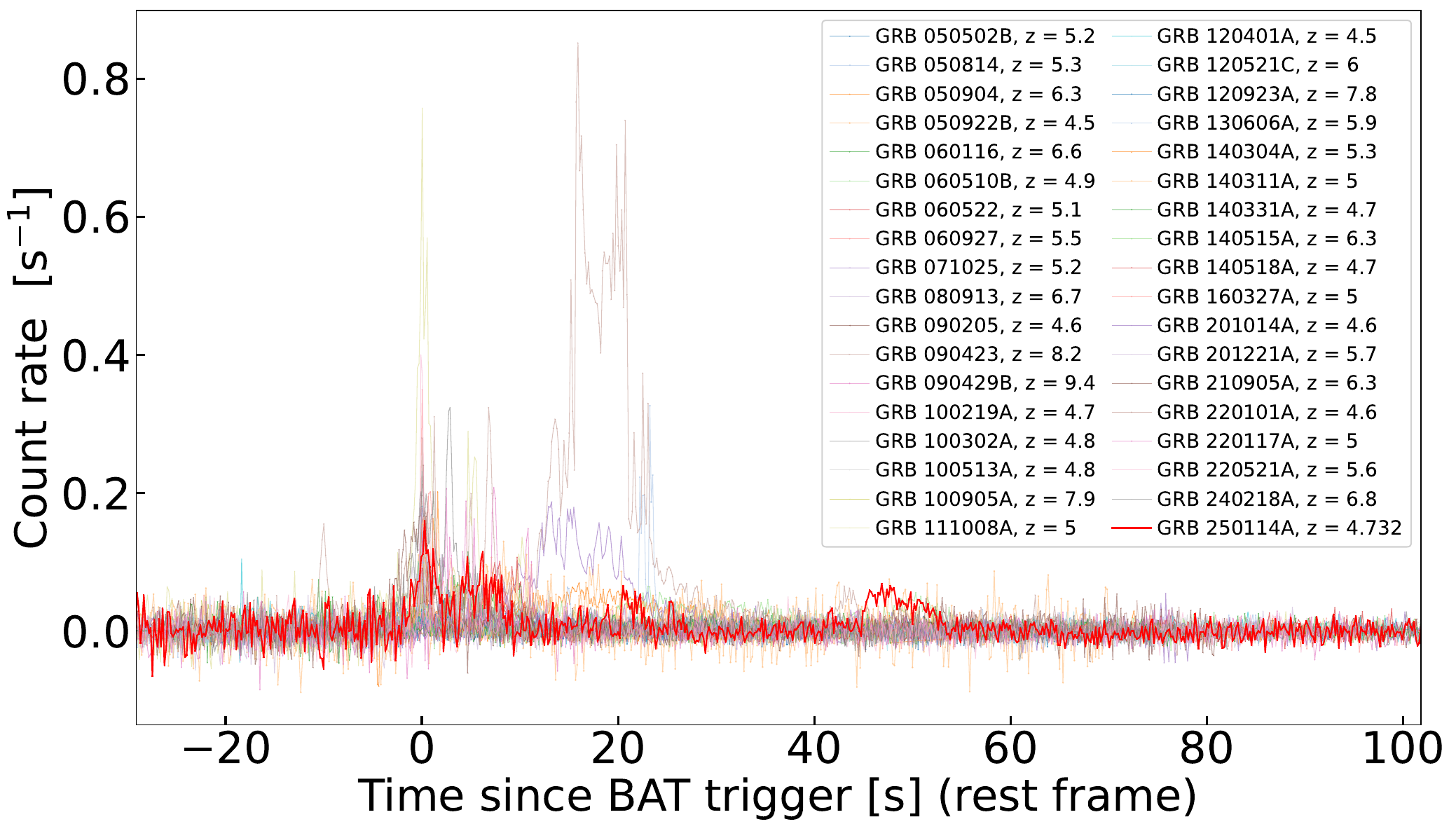}
\caption{Comparison of GRB 250114A and Swift sample-GRBs with $z \textgreater 4.5$ in prompt emission.
\label{fig3}}
\end{figure}

Over the past few decades, Swift has detected hundreds of GRBs with measured redshifts, yet those at $z>4.5$ remain a tiny minority. The rate of GRBs detected by Swift which end up to have their redshift measured with $z>4.5$ is approximately 1.5 events per year. In order to compare the properties of prompt emission and X-ray afterglow of GRB 250114A and those of other high-$z$ (e.g., $z>$4.5) GRBs, we first select the sample-GRBs which are taken from \citep{2025NatAs...9..564L}, and then extract the light curves of both BAT and XRT. Figure~\ref{fig3} shows the prompt emission light curves of sample-GRBs and GRB 250114A in the rest frame. From a statistical perspective, the duration of GRB 250114A in the rest frame appears to be longer than that of other high-$z$ sample-GRBs. Moreover, we also plot the X-ray afterglow of GRB 250114A together with those of high-$z$ sample-GRBs in the rest frame, we find that no statistically significant difference between them.

On the other hand, \cite{2002A&A...390...81A} discovered that a majority of long-duration GRBs exhibit a positive correlation between the peak energy ($E_{\rm p}$) and the isotropic burst energy ($E_{\rm \gamma, iso}$), namely, $E_{\mathfrak{p}}\propto E_{\gamma,\mathrm{iso}}^{1/2}$, even the dispersion of the correlation is large, and outliers do exist \citep{2009ApJ...703.1696Z}. However, the Amati relation of the majority of short-duration GRBs appears to be a slightly shallower in terms of power index when compared with that of long GRBs \citep{2009ApJ...703.1696Z}. \cite{2010ApJ...725.1965L} proposed a new phenomenological classification method for GRBs by adopting a new parameter $\varepsilon = E_{\rm \gamma,iso,52}/E^{\rm 5/3}_{\rm p,z,2}$, which has proven to be highly effective in the classification of high-$z$ GRBs. Here, $E_{\rm \gamma,iso,52}$ is the isotropic gamma-ray energy in units of $10^{52}$ erg, and $E_{\rm p,z,2}$ is the cosmic rest-frame spectral peak energy in units of 100 keV. They found that the $\varepsilon$ has a clear bimodal distribution (high- and low- $\varepsilon$ regions) with a division line at $\varepsilon \sim 0.03$, and the low- and high-$\varepsilon$ regions corresponding to the mergers of two compact stars (type I) and the death of massive stars (type II), respectively.

By calculating the isotropic burst energy via k-correction \citep{2001AJ....121.2879B}, one has $E_{\gamma, \mathrm {iso}} = (\mathrm{1.7} \pm 0.2) \times10^{53}\,\mathrm{erg}$ for GRB 250114A. Due to the lack of enough optical afterglow data, it is difficult to directly constrain the jet opening angle for GRB 250114A. Alternatively, we adopt the empirical Ghirlanda relation which presented the relationship between the jet opening angle, the isotropic gamma-ray energy, and the peak energy of the burst spectrum \citep{2016ApJ...818...18G} to estimate the opening angle. If this is the case, one can roughly derive a jet opening angle as $\theta_{\rm j}=3.14^{\circ}$ (corresponding to a beaming factor $f_{\mathrm b} = 1- \rm {cos}(\theta_{\rm j})\sim 0.0015$), and the jet corrected energy is about $E_{\gamma} \sim \mathrm{2.55} \times10^{50}\,\mathrm{erg}$. 

Figure~\ref{fig5}(a) shows the $E_{\rm p}-E_{\rm \gamma, iso}$ correlation for GRB 250114A and other long-duration GRBs. It is found that GRB 250114A falls squarely on the main sequence of long-duration GRBs, thereby suggesting the potential for a shared progenitor with that of other long-duration GRBs. Adopting the method from \cite{2010ApJ...725.1965L}, one can calculate $\varepsilon \sim 2.84$, which is located in the high-$\varepsilon$ region. The high-$\varepsilon$ value of GRB 250114A suggests that its progenitor is consistent with Type II population from the death of massive stars (see Figure~\ref{fig5}). Furthermore, wo also plot the high-$z$ GRBs in the $E_{\rm p}-E_{\rm \gamma, iso}$ and $\varepsilon-$distribution diagrams to facilitate a comparison with GRB 250114A, as illustrated in Figure~\ref{fig5}. It has been established that the location of most high-$z$ GRBs is situated within into the main sequence for long-duration GRBs in the $E_{\rm p}-E_{\rm \gamma, iso}$ diagram and the high-$\varepsilon$ region.

On the other hand, the $E_{\rm p}-E_{\rm \gamma, iso}$ diagram is not foolproof method to identify the progenitors \citep{2025A&A...698A.250K}. Alternatively, another well-established empirical correlation is so-called Dainotti relation \citep{2013MNRAS.436...82D, 2014MNRAS.443.1779R} which is used to judge the origin of core-collapse for GRBs. This relation reveals an anti-correlation between the duration of X-ray plateau in rest-frame  ($T_a^*$) and the X-ray plateau luminosity ($L_X$) of GRBs, which has been confirmed in large samples of long-duration bursts and proposed as a potential tool for GRB standardization \citep{2025JHEAp..4700384L}. For GRB~250114A, one can obtain $T_a^* = (2.76 ^{+0.40}_{-0.40})\times 10^3$~s and $L_X = (5.63 ^{+0.69}_{-0.62} )\times 10^{47}$~erg~s$^{-1}$ from our XRT afterglow fits, and it is found that it is also located within the core-collapse (Type~II) region.

\section{Physical Interpretation}\label{sec:model}
The long‐duration prompt emission of GRB 250114A suggests that the progenitor is likely related to the collapse of a massive star. In the collapsar scenario, a hyper‐accreting black hole or a rapidly spinning magnetar may be formed as the central engine \citep{1992Natur.357..472U, 1998PhRvL..81.4301D, 2001ApJ...552L..35Z,2013ApJ...765..125L, 2017NewAR..79....1L}. It is imperative that the engine remains operational for a sufficient duration to ensure the jet head successfully breaks out of the star and produces a successful jet. Based on the feature of X-ray afterglow emission of GRB 250114A, a magnetar with a high surface magnetic field and a short-period is proposed to be a central engine of GRB 250114A. It is evident that the observed X-ray plateau emission is attributable to magnetic dipole radiation from the newborn millisecond magnetar prior to its collapse \citep{1998PhRvL..81.4301D, 2001ApJ...552L..35Z, 2011MNRAS.413.2031M, 2014ApJ...785...74L}.

The total rotational energy of the newborn millisecond magnetar can be expressed as \citep{2001ApJ...552L..35Z, 2018MNRAS.480.4402L}:
\begin{equation}
    E_{\mathrm{rot}} = \frac{1}{2}I\Omega^{2} \approx 2 \times 10^{52} \, \mathrm{erg} \, M_{1.4}R_{6}^{2}P_{0,-3}^{-2},
\end{equation}
where $I$ is the moment of inertia, $\Omega$ is the angular frequency, $P_{0,-3}$ is the initial period of the neutron star in units of milliseconds, $R_6$ is the radius of the neutron star in units of $10^6$ cm, $M_{1.4} = 1.4 M_{\odot}$ is the mass of neutron star. By considering a newly born magnetar, it is spun down via electromagnetic (EM) dipole radiation,
\begin{eqnarray}
-\frac{dE_{\rm rot}}{dt} = -I\Omega \dot{\Omega} = L_{\rm EM}
= \frac{B^2_{\rm p}R^{6}\Omega^{4}}{6c^{3}},
\label{Spindown}
\end{eqnarray}
\begin{eqnarray}
L_{\rm EM}(t) &=& L_{0}(1+\frac{t}{\tau})^{-2}.
\label{Luminosity_EM}
\end{eqnarray}
Here, $B_{\rm p}$ is surface magnetic filed, $c$ is speed of light, $L_{0}$ and $\tau$ are the initial kinetic luminosity and the characteristic spin-down timescales, respectively, namely,
\begin{eqnarray}
L_{0}&=&1.0 \times 10^{49}~{\rm erg~s^{-1}} (B_{\rm p,15}^2 P_{0,-3}^{-4} R_6^6),
\label{spinlu_em}
\end{eqnarray}
\begin{eqnarray}
\tau= 2.05 \times 10^3~{\rm s}~ (I_{45} B_{\rm p,15}^{-2} P_{0,-3}^2 R_6^{-6}),
\label{spintau_em}
\end{eqnarray}
where $B_{\rm p,15}$ is the magnetic filed strength in units of $10^{15}$ $\rm G$, and $I_{45}$ is the moment of inertia in units of $10^{45} \ \rm g \ cm^2$.

\subsection{Model fitting with magnetar central engine}
In this section, we invoke the with the \texttt{full vacuum dipole magnetar} ~\citep{2001ApJ...552L..35Z, 2013MNRAS.430.1061R} central engine model to fit the flare-subtracted X-ray plateau emission\footnote{We excluded X-ray flare emission which is believed to be from central engine re-activity in our model fitting.} by adopting the open-source software package \texttt{Redback}~\citep{2024MNRAS.531.1203S}. Since such magnetar model is also employed in the context of numerous long-duration GRBs exhibiting X-ray plateau emission \citep{2006MNRAS.372L..19F, 2007ApJ...665..599T, 2010MNRAS.402..705L, 2012A&A...539A...3B, 2024RAA....24h5003S}. Although recent work has explored the possible role of higher-order multipolar fields (e.g., quadrupole, octupole) in shaping the early-time light-curve behavior \citep{2024ApJ...974...89W}, such effects require rather extreme initial configurations. Based on the fitting results, the plateau luminosity of GRB 250114A is about $2\times 10^{48}\rm ~erg/s$ which is consistent with the numerical calculation with dipole radiation dominated \citep{2024ApJ...974...89W}. On the other hand, the decay slope after the plateau is steeper than that predicted from the higher-order field components. So that, we only consider the dipole radiation of magnetar in this work to do the model fitting.

In the magnetar framework, the total luminosity is modeled as the sum of a dipole spin‐down component and an underlying power‐law decay,
\begin{equation}
    L(t)=L_{0}\!\left(1+\frac{t}{\tau}\right)^{\frac{1+n}{1-n}} \;+\; A_{1}\,t^{-\alpha_{1}}\,,
\end{equation}
where \(n\) is the braking index of magnetar \citep{2017ApJ...843L...1L}. The extra power‐law term with amplitude \(A_{1}\) and index \(\alpha_{1}\) accounts for the high‐latitude (curvature) emission of the prompt phase \citep{2000ApJ...541L..51K}. The time of the observer frame is converted to $t/(1+z)$ in the source frame, and the bolometric luminosity \(L(t)\) in the above equation is derived from the observed X-ray flux of plateau \(F_{\mathrm{obs}}\) directly into an isotropic bolometric luminosity,
\begin{equation}
L (t) = 4\pi D_{L}^{2} F_{\mathrm{obs}},
\end{equation}
where \(D_{L}\) is the luminosity distance.

There are five free parameters (i.e., $L_{\mathrm 0}$, $\tau$, $n$, $A_{\mathrm 1}$ and $\alpha_{\mathrm 1}$) by invoking the \texttt{full vacuum dipole magnetar} model. The prior distributions for the parameters were defined as follows:
\[
\left\{
\begin{aligned}
A_{1} &\sim \mathrm{LogUniform}\bigl(10^{-20},\,10^{20}\bigr)\,,\\
\alpha_{1} &\sim \mathrm{Uniform}\bigl(-10,\,-0.5\bigr)\,,\\
L_{0} &\sim \mathrm{LogUniform}\bigl(10^{30},\,10^{55}\bigr)\,\mathrm{erg\,s^{-1}},\\
\tau &\sim \mathrm{LogUniform}\bigl(10^{2},\,10^{6}\bigr)\,\mathrm{s},\\
n &\sim \mathrm{Uniform}\bigl(1,\,7\bigr)\,.
\end{aligned}
\right.
\]
To sample the posterior distributions, we employ the \texttt{dynesty} sampler \citep{2020MNRAS.493.3132S}, a pure Python, MIT‑licensed dynamic nested sampling package for estimating Bayesian posteriors and evidences. We configure \texttt{dynesty} with \texttt{nlive=1000} live points and use the \texttt{sample='rslice'} method. The \texttt{rslice} option implements random‑direction slice sampling: at each iteration, a random direction in the parameter space is chosen and a one‑dimensional slice sampling step is performed along that direction, which improves mixing and exploration efficiency in multi‑dimensional posteriors.

\subsection{Fitting Results and inferred Magnetar Parameters}

Figure \ref{fig6} shows the fitting results by using the magnetar model in the rest-frame to the XRT light curve with flare-subtracted of GRB 250114A. The 2D histograms and parameter constraints of the model fit by invoking the \texttt{dynesty} sampler are shown in Figure \ref{fig6}, and the values of the parameter constraints are reported in the Table \ref{table3}. It is found that four free parameters (i.e., $L_{\mathrm 0}$, $n$, $A_{\mathrm 1}$ and $\alpha_{\mathrm 1}$), can be constrained very well. Based on the results of best‐fit for the spin‐down luminosity $L_{\mathrm 0} =2.78^{+0.46}_{-0.42} \,\times 10^{48}\, \mathrm{erg \, s^{-1}}$ and characteristic timescale $\tau= 2397^{+1057}_{-310} \, \mathrm s$, one can directly infer the surface polar cap magnetic field strength $B_{\mathrm p}$ and initial spin period $P_{\mathrm 0}$ of the magnetar by adopting Eqs. (8) and (9).

It is worth noting that a gap in the X-ray light curve between $t\sim2\times10^{2}\,\mathrm{s}$ and $t\sim8\times10^{2}\,\mathrm{s}$ in the rest frame should be a potential systematic uncertainty in the identification of the magnetar spin-down plateau. In particular, the emission at $t\sim10^{2}\,\mathrm{s}$ is possible to be contributed from both tail emission of the X-ray flare and spin-down of magnetar. In order to confirm the validity of this hypothesis, the fits were repeated using data exclusively from $t\geq8\times10^{2}\,\mathrm{s}$ (i.e., excluding the flare and power-law components). It is found that the spin-down timescale and the braking index $n$ become poorly constrained with larger uncertainties. Therefore, although we believe the possibility that flare contamination may affect the earliest post-gap points, the flare-subtracted fit provides the most robust and tightly constrained interpretation of the available data.

By adopting the derived a jet opening angle of $\theta_{\rm j}=3.14^{\circ}$ from \citealt{2016ApJ...818...18G} and assuming a radiation efficiency $\eta\sim 0.1$, one can estimate the magnetar initial spin period and polar magnetic field strength to be $P_{0}=14.31^{+0.93}_{-3.16} \,  \mathrm{ms}$ and 
$B_{\rm p}=13.24^{+1.73}_{-5.84} \, \times10^{15}\ \mathrm{G}$, respectively. Moreover, by comparing the inferred $B_{\mathrm p}$ and $P_{\mathrm 0}$ of GRB 250114A with that of other magnetar candidates GRBs taken from \cite{2014ApJ...785...74L}, we find that the derived values of those parameters fall into the reasonable range (see Figure \ref{fig7}). 

Moreover, the early peak of optical emission is likely to be associated with the prompt emission of GRB itself, as shown in Figure \ref{fig2}. While the lack of enough optical observations at late time seems to be a peak or a plateau emission which is consistent with X-ray emission, but it is difficult to confirm due to the few optical observations.

\section{Conclusions} \label{sec:conclusion}
GRB 250114A is a long-duration GRB that triggered Swift/BAT \citep{2025GCN.38927....1K}, and the spectroscopic observations by the Very Large Telescope derived a redshift at $z$ = 4.732 \citep{2025GCN.38934....1M}. The prompt emission of GRB 250114A shows three distinct emission episodes which are identified via the Bayesian Blocks algorithm \citep{1998ApJ...504..405S}. The X-ray light curve displays the canonical X-ray emission that appears to be composed of multiple power-law segments: an initial steep decay smoothly connected with a plateau phase and followed by a normal decay segment, and a giant X-ray flare is superimposed on the steep decay segment. More interestingly, it is found that the X-ray flare emission of GRB 250114A was simultaneously observed by BAT in episode- \uppercase\expandafter{\romannumeral2} and episode-\uppercase\expandafter{\romannumeral3} by comparing the light curves of X-ray and $\gamma-$ray. There is still significant X-ray emission during the quiescent period between the episode-\uppercase\expandafter{\romannumeral2} and episode-\uppercase\expandafter{\romannumeral3}.

By calculating the isotropic burst energy and extracting the spectra of GRB 250114A, one has $E_{\gamma, \mathrm {iso}} = (\mathrm{1.7} \pm 0.2) \times10^{53}\,\mathrm{erg}$ and $E_{\mathrm{p}} = 51\pm 6$ keV. GRB 250114A is located at the main sequence of long-duration GRBs in the $E_{\rm p}-E_{\rm \gamma, iso}$ diagram, and is consistent with that of other long GRBs in high-$\varepsilon$ region. It suggests that GRB 250114A is possible shared a similar progenitor from the death of massive stars with that of other long-duration GRBs. Moreover, by comparing both prompt emission and X-ray afterglow with that of other high-$z$ GRBs, it is found that the duration of GRB 250114A seems to be longer than that of other high-$z$ sample-GRBs, but the peak flux of pulse of GRB 250114A is mixing with that of other high-$z$ sample-GRBs. The X-ray afterglow appears to demonstrate no statistically significant difference between between them. In the $E_{\rm p}-E_{\rm \gamma, iso}$ diagram and $\varepsilon-$distribution, the location of most high-$z$ GRBs fall into the main sequence for long-duration GRBs in $E_{\rm p}-E_{\rm \gamma, iso}$ and high-$\varepsilon$ region, respectively. Moreover, we find that GRB 250114A is also located within the core-collapse (Type~II) region of Dainotti relation \citep{2010ApJ...722L.215D,2011MNRAS.418.2202D,2013MNRAS.436...82D,2014MNRAS.443.1779R}, and the plateau luminosity-duration anti-correlation is also confirmed by other researchers in the literature \citep{2018ApJ...869..155S}.

Based on the feature of X-ray afterglow emission of GRB 250114A, we propose a magnetar as a central engine of GRB 250114A. The observed X-ray plateau emission is hypothesised to be attributable to  magnetic dipole radiation from the newborn millisecond magnetar prior to its collapse. Therefore, we model the X‐ray plateau using the full vacuum dipole magnetar scenario implemented in \texttt{Redback} and sampled with \texttt{dynesty}. The magnetar central engine model provides well fit to the flare-subtracted light curve, yielding well‐constrained posterior distributions for $L_0$, $\tau$, $n$, $A_1$, and $\alpha_1$. If this is the case, we also infer the magnetic field strength $B_{\rm p}=13.24^{+1.73}_{-5.84} \, \times10^{15}\ \mathrm{G}$ and initial spin period $P_{0}=14.31^{+0.93}_{-3.16} \,  \mathrm{ms}$ with a jet correction, and it also falls into the reasonable range by comparing other candidates of magnetar central engine in long-duration GRBs.   

Within the magnetar central engine scenario, the quiescence period (e.g., three episodes emission) in the prompt phase is possibly caused by accretor-propeller behaviour of the magnetar which is proposed by \citealt{2013ApJ...775...67B}. The $\gamma-$ray emission of episodes I and II are likely powered by active accretion onto the newborn magnetar, and the magnetic field rises rapidly and temporarily repels the infalling material. Therefore, the electromagnetic emission temporarily fades. The subsequent quiescent period in $\gamma-$ray, accompanied by significant X-ray emission, could mark a transition from accretor-propeller to the propeller regime. During this phase, the rapidly spinning magnetar's intense magnetic field effectively repels the infalling material, halting efficient accretion onto the surface and thus quenching the high-energy $\gamma-$ray production. However, the central engine remains active, the rotational energy of the magnetar can be dissipated through magnetic braking or interactions with the halted material, giving rise to the observed lower-energy X-ray emission. The eventual restart of intense $\gamma-$ray emission in episode-III would then correspond to the central engine overcoming this barrier and resuming active accretion. The observed evolution of the spectral peak energy $E_p$ is also consistent with this picture of intermittent energy injection from a magnetar.

\section{Discussion} \label{sec:Discussion}
If the magnetar is indeed operating as the central engine in GRB 250114A, one question is that whether the total energy observed from GRB 250114A satisfies the energy budget of the newborn magnetar. The plateau energy can be calculated as $E_{\rm X,iso}\sim L_{x}\times T_a^* \sim 1.56\times 10^{51}$ erg, and the kinetic energy can be roughly estimated as $E_{\rm K,iso} \sim 2.4\times 10^{54}$ erg by adopting the method in \citealt{2007ApJ...655..989Z} and \citealt{2018ApJS..236...26L}. We find that the total energy of magnetar is larger than that sum of $\gamma-$ray energy, plateau energy, and kinetic energy after jet correction, namely, $E_{\rm tot}>(E_{\rm \gamma,iso}+E_{\rm X,iso}+E_{\rm K,iso})f_{\rm b}$. It suggests that at least the energy budget of magnetar central engine satisfies with output energy of observations. Excepting magnetar central engine model, the black hole central engine as a candidate to interpret GRB 250114A can not be ruled out. For example, fallback accretion onto a black hole \citep{2008MNRAS.388.1729K,2009ApJ...700.1047C,2011ApJ...734...35C,2017ApJ...849...47L,2025JHEAp..4700384L}, or continued central engine activity from the hydrodynamical interactions of a single blast shock wave with the circumburst medium (e.g., \citealt{2020ApJ...900..193D,2022NatCo..13.5611D}).

Moreover, the physical explanation of GRB 250114A that we adopt may not be the only one to the observations, and several physical models have also been invoked to interpret the observed X-ray flare, plateau emission, as well as the multi-episode prompt emission. For example, \cite{2006Sci...311.1127D} propose that the differential rotation of millisecond pulsar winds up internal poloidal fields into strong toroidal fields that buoyantly break through the surface and reconnect explosively, powering the X‑ray flare. \cite{2017ApJ...834...28N} have shown that when the GRB jet breaks out of the star it inflates a cocoon whose expansion and interaction with the external medium can power wide‑angle X‑ray flares, with their luminosity and duration set by the cocoon’s energy deposition and the degree of jet–stellar mixing. \cite{2006ApJ...636L..29P} and \cite{2017MNRAS.464.4399D} also proposed that fragmentation in the accretion disk can reproduce the empirical statistical properties of the X-ray flares. \cite{2023ApJ...945...17G} propose a jet precession model to interpret the repeating emission episodes. \cite{2007ApJ...670.1247W} proposed the jet bow shock and relativistic jet itself to explain the two-episode emission. 

Furthermore, other possible interpretations are also proposed, such as fallback accretion on the magnetar \citep{2024ApJ...962....6Y}, or the magnetic barrier around the accretor \citep{2006MNRAS.370L..61P}, or the gravitational lensing \citep{2021NatAs...5..560P,2021ApJ...921L..29Y}. However, more or less, each model above cannot fully interpret all properties of the observations of GRB 250114A from the light curve, spectrum, as well as the quiescent time. The magnetar model currently presents the most self-consistent framework for simultaneously explaining the sustained activity across energy bands and the specific temporal evolution observed in this event.

One is needed to note that the connection between long GRBs and the cosmic star formation rate (SFR) remains debated. Although some studies suggest that long GRBs broadly trace the SFR \citep{2015A&A...581A.102V}, other works have reported deviations \citep{2022A&A...666A..14S,2023ApJ...958...37D}. These discrepancies may in part be explained by the existence of a distinct population of low-luminosity GRBs \citep{2009MNRAS.392...91V}, or by possible contamination of the long-GRB sample by long-duration compact binary mergers \citep{2024ApJ...963L..12P}. More recent analyses also indicate that the GRB–SFR relation may be more complex than previously assumed, requiring consideration of multiple populations and evolutionary effects \citep{2018A&A...610A..58A, 2022ApJ...933...17A, 2024ApJ...976L..16M}.

On the other hand, the nature of the significant X-ray emission during the quiescent period remains in debate. Although most studies shown that the interval between pulses of prompt emission is a true “quiescent period”. It indicates that the central engine switches off during such quiescent period, subsequently re-initiating to power the sub-bursts of gamma-ray emission and X-ray flares in the afterglow. However, at least for GRB 250114A, it is found that the central engine never truly shuts down; rather, it persists in its activity. Just because of the $\gtrsim10\,$keV $\gamma$‑rays emission are absence, but is replaced by lower-energy X-ray emission. In summary, the central engine of GRB 250114A has been in operation to power those three-episode emissions and X-ray flare, even during the quiescent period in $\gamma$‑rays emission.

\section{Acknowledgments}
We acknowledge the use of the public data from the Swift/BAT and XRT Science Data Center. This work is supported by the Natural Science Foundation of China (grant Nos. 12494574, 11922301, and 12133003), the Natural Science Foundation of Guangxi (grant Nos. 2023GXNSFDA026007 and 2025FNFN99005), the Program of Bagui Scholars Program (LHJ), and the Guangxi Talent Program (Highland of Innovation Talents).


\begin{table}[ht]
\caption{The Fitting Results of Time-resolved Spectra with CPL model for GRB 250114A}
\centering
\begin{tabular}{cccccccc}
\toprule
Time interval (s) & -10-20 &20-50 &112-137 & 210-232 & 232-253 & 253-282 & 282-320 \\
\midrule
$\Gamma$ & $-0.02 \pm 0.73$ & $0.30 \pm 0.62$ & $1.23 \pm 0.02$ & $0.96 \pm 0.17$ & $1.06 \pm 0.06$ & $0.72 \pm 0.03$ & $0.91 \pm 0.03$  \\
\midrule
$E_\mathrm{p}$ (keV) & $23 \pm 10$ & $52 \pm 31$ & $55 \pm 16$ & $7 \pm 4$ & $28 \pm 13$ & $46 \pm 5$ & $50 \pm 11$  \\
\bottomrule
\label{table1}
\end{tabular}
\end{table}

\begin{table}[ht]
\caption{The empirical Fitting Results of the X-ray Afterglow Light Curves with power-law $+$ two smooth broken power-law models for GRB 250114A}
\centering
\begin{tabular}{cccc}
\toprule
 & \makecell{$\alpha$} &  \makecell{$t_b$\\(s)} &  \makecell{$F_b$\\(erg cm$^{-2}$ s$^{-1}$)} \\
\midrule
\makecell{Powe-law} & \makecell{$\alpha_1 = 4.22_{-0.18}^{+0.19}$} & \makecell{-} & \makecell{-} \\
\midrule
\makecell{Flare} & \makecell{$\alpha_2 = -9.14_{-0.38}^{+0.38}$ \\ $\alpha_3 = 7.52_{-0.14}^{+0.15}$} & \makecell{$t_{\rm b,2} \sim 278 $} & \makecell{$F_{\rm b,2} = (4.33\pm 0.08) \times 10^{-9}$} \\
\midrule
\makecell{Plateau} & \makecell{$\alpha_4 = 0.53_{-0.03}^{+0.02}$ \\ $\alpha_5 = 1.70_{-0.28}^{+0.38}$} & \makecell{$t_{\rm b,3} = 15859_{-2343}^{+2531}$} & \makecell{$F_{\rm b,3} = (2.48\pm 0.30) \times 10^{-12}$} \\
\bottomrule
\label{table2}
\end{tabular}
\end{table}

\begin{table}[ht]
\caption{The Parameters of magnetar model fits and its inferred for GRB 250114A}
\centering
\begin{tabular}{cccccccc}
\toprule
Parameter & $A_{\mathrm 1}$ &$\alpha_{\mathrm 1}$ & $L_{\mathrm 0} \, (10^{48}\, \mathrm{erg \, s^{-1}})$ & $\tau \, (\mathrm s)$ & $n$ & $B_{\mathrm{p}} \, (\times10^{15}\ \mathrm{G})$ & $P_{\mathrm{0}} \,(\mathrm{ms})$\\
\midrule
Value & $117525^{+52577}_{-34038}$ & $-3.22^{+0.10}_{-0.11}$ & $2.78^{+0.46}_{-0.42}$ & $2397^{+1057}_{-310}$ & $2.44^{+0.55}_{-0.45}$ & $13.24^{+1.73}_{-5.84}$& $14.31^{+0.93}_{-3.16}$\\
\bottomrule
\label{table3}
\end{tabular}
\end{table}

\begin{figure}[htbp]
\centering
\includegraphics[width=0.6\textwidth]{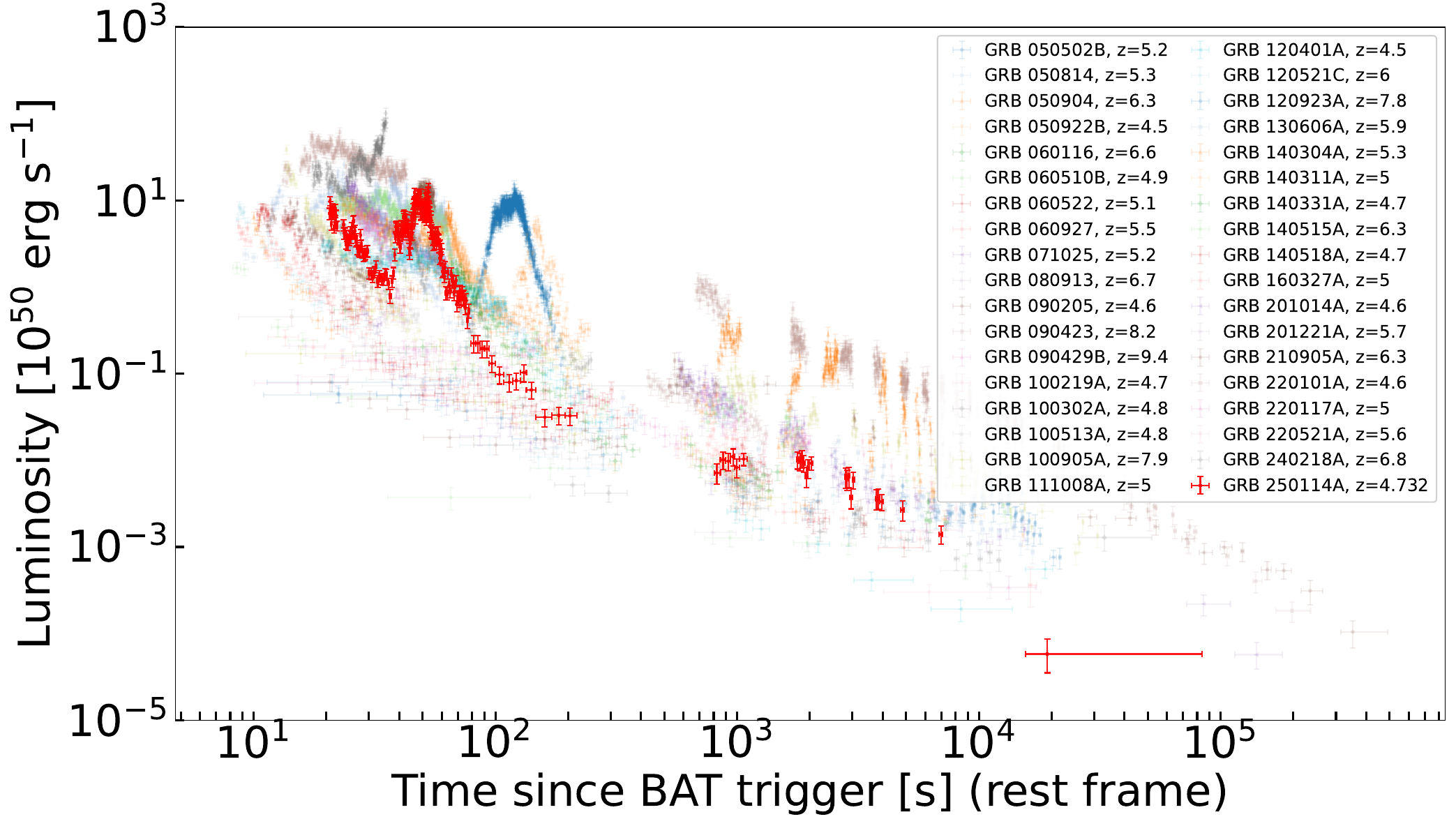}
\caption{Comparison of GRB 250114A and Swift sample-GRBs with $z \textgreater 4.5$ in X-ray afterglow.
\label{fig4}}
\end{figure}


\begin{figure}[htbp]
\centering
\begin{minipage}[c]{0.53\textwidth} 
  \centering
  \includegraphics[width=\linewidth]{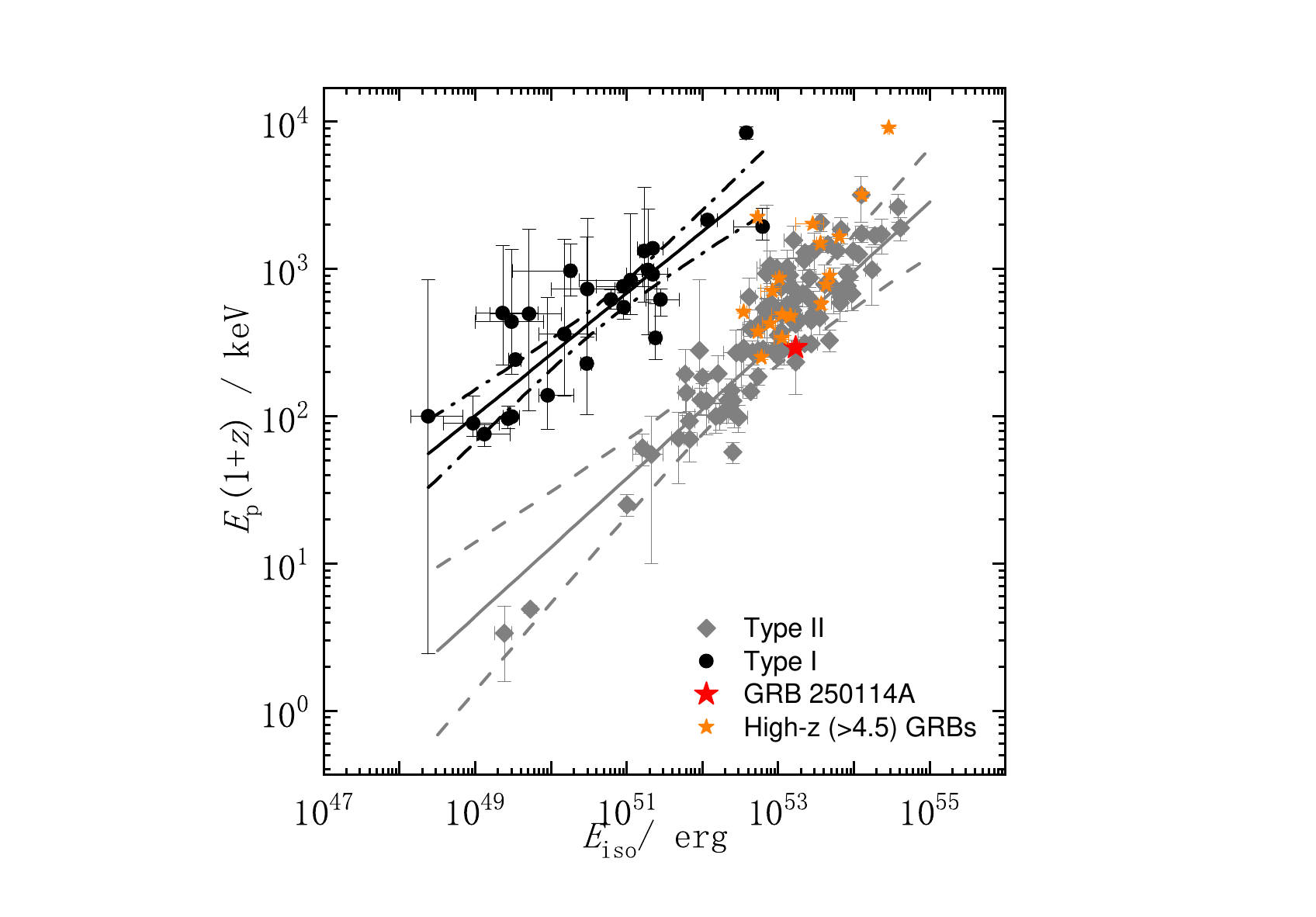} 
\end{minipage}
\hfill
\begin{minipage}[c]{0.44\textwidth}
  \centering
  \includegraphics[width=\linewidth]{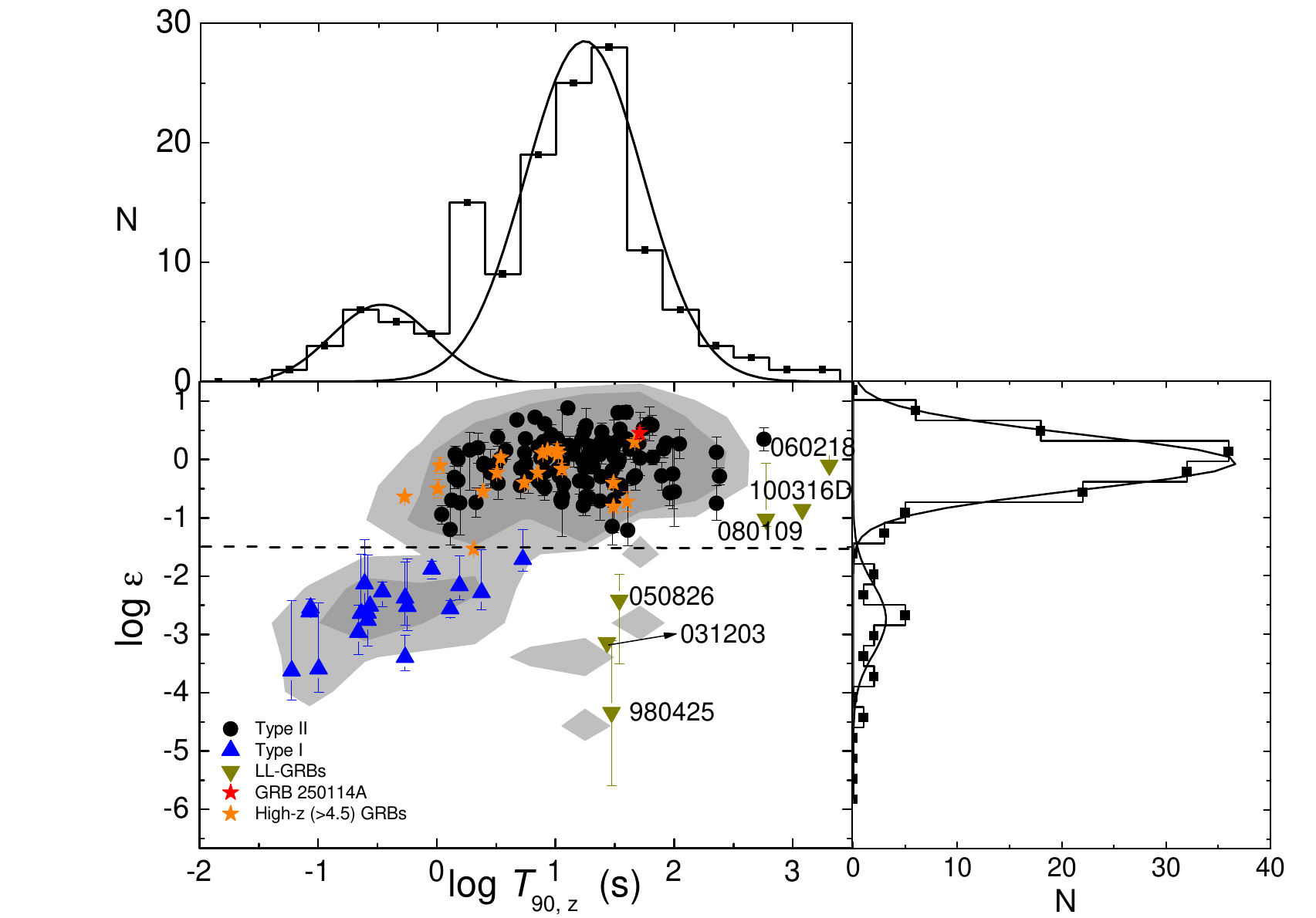}

\end{minipage}
\caption{Left panel: $E_{\mathrm p}$ and $E_{\gamma, \mathrm {iso}}$ correlation diagram. Black points and gray diamonds correspond to Type I and Type II GRBs, respectively. The red star is GRB 250114A, and other data are taken from \cite{2009ApJ...703.1696Z}. Right panel: 1D and 2D distributions of GRB samples in $T_{\mathrm{90}} - \epsilon$ space. The dashed line indicates $\epsilon = 0.03$, and the data come from \cite{2010ApJ...725.1965L}}
\label{fig5}
\end{figure}

\begin{figure}[htbp]
\centering
\begin{minipage}[b]{0.5\textwidth} 
  \centering
  \includegraphics[width=\linewidth]{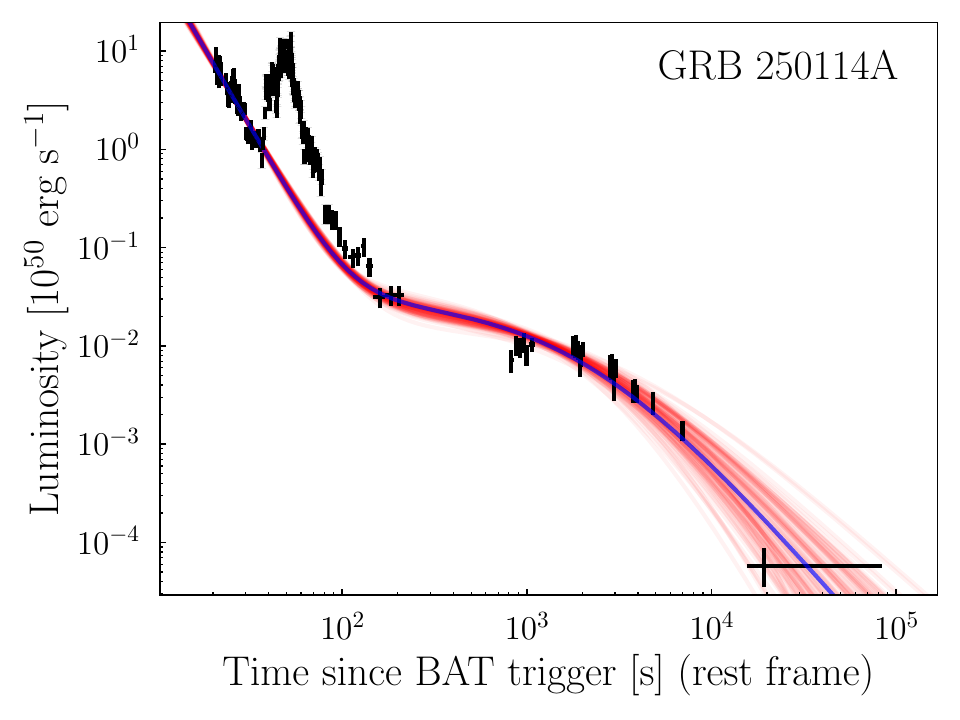} 
\end{minipage}
\hfill
\begin{minipage}[b]{0.45\textwidth}
  \centering
  \includegraphics[width=\linewidth]{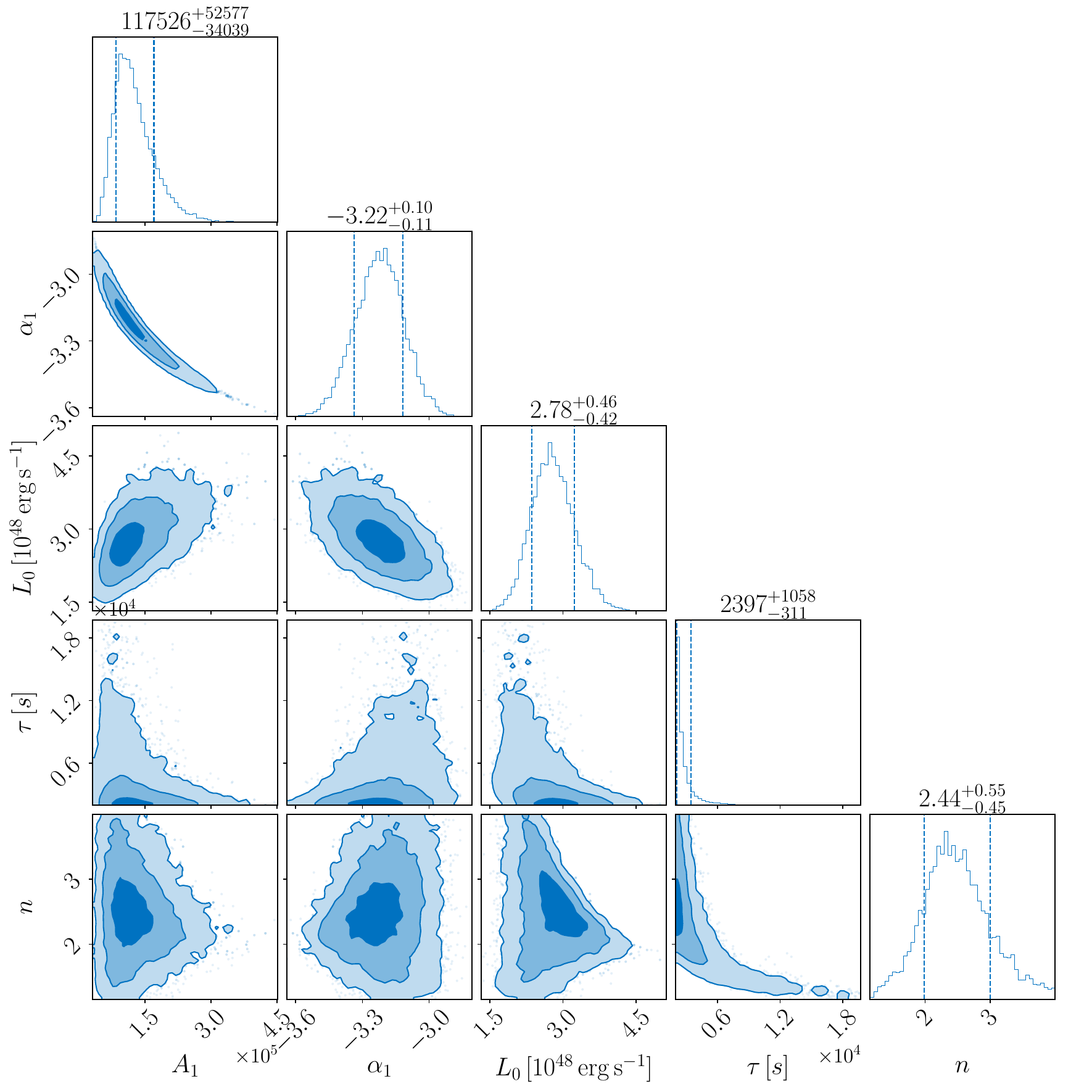}

\end{minipage}
\caption{Left panel: X-ray light curve of GRB 250114A (black points) with magnetar model fits. The solid black line represents the maximum likelihood model, while the red lines represent 100 posterior samples, illustrating the uncertainty in the model fit. Right panel: 2D histograms and parameter constraints of the model fits for GRB 250114A. The 1D histograms show the distributions for each parameter. The dashed lines indicate the median and $\pm 2 \sigma$ uncertainty of the values.}
\label{fig6}
\end{figure}


\begin{figure}[h]
\centering
\includegraphics[width=0.9\textwidth]{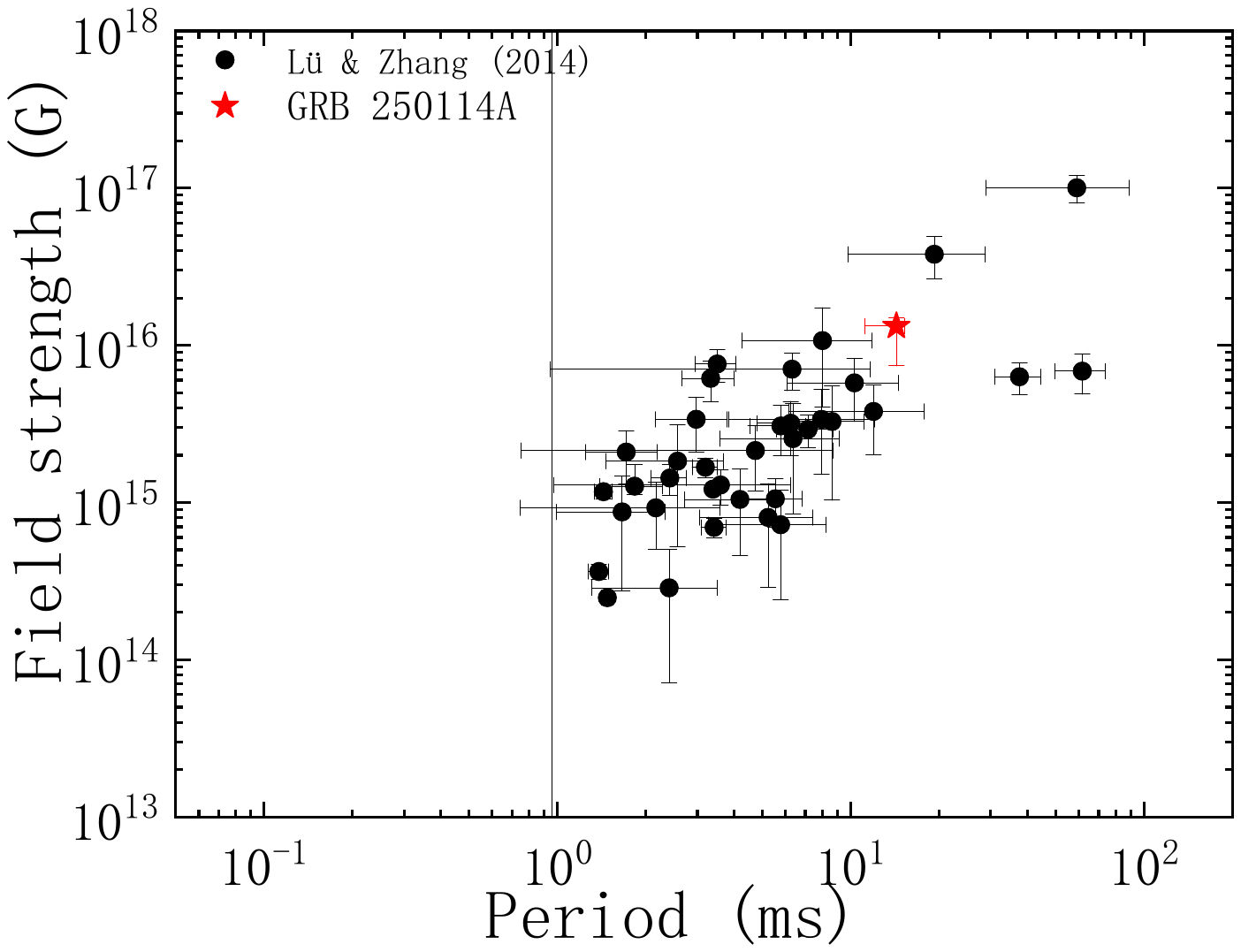}
\caption{Inferred magnetar parameters of GRB 250114A, initial spin period $P_{\mathrm 0}$ vs. surface polar cap magnetic field strength $B_{\mathrm{p}}$. Other candidates for the magnetar cental engine are taken form \cite{2014ApJ...785...74L}. The vertical solid line is the breakup spin period for a neutron star \citep{2004Sci...304..536L}.
\label{fig7}}
\end{figure}

\end{document}